# Comprehensive and Reliable Crowd Assessment Algorithms

Manas Joglekar
Stanford University
Stanford, California 94305
Email: manasrj@stanford.edu

Hector Garcia-Molina
Stanford University
Stanford, California 94305
Email: hector@cs.stanford.edu

Aditya Parameswaran
University of Illinois at Urbana Champaign
Champaign, Illinois 61801
Email: adityagp@cs.stanford.edu

*Abstract*—Evaluating workers is a critical aspect of any crowdsourcing system. In this paper, we devise techniques for evaluating workers by finding confidence intervals on their error rates. Unlike prior work, we focus on "conciseness"—that is, giving as tight a confidence interval as possible. Conciseness is of utmost importance because it allows us to be sure that we have the best guarantee possible on worker error rate. Also unlike prior work, we provide techniques that work under very general scenarios, such as when not all workers have attempted every task (a fairly common scenario in practice), when tasks have non-boolean responses, and when workers have different biases for positive and negative tasks. We demonstrate conciseness as well as accuracy of our confidence intervals by testing them on a variety of conditions and multiple real-world datasets.

## I. Introduction

A crowdsourcing system employs human workers to perform data processing tasks, typically on unstructured data, such as filtering and comparisons, which humans can perform better than computers can. An important issue in such a system is the abilities of workers who are performing tasks. Most workers make some mistakes while working, but different workers may have different rates of making mistakes while performing tasks. This makes it important to estimate the error rate of each worker so that workers with high error rates can be replaced or trained, while workers with low error rates are retained for future tasks.

The classical way to evaluate workers is to have workers execute a number of *gold standard* tasks for which correct responses are known in advance. By comparing the worker responses with the correct responses, we can use standard statistical techniques [1] to compute confidence intervals. However, it is often problematic and expensive to develop gold standard tasks. First, expert workers must be paid to identify the correct responses. Moreover, the gold standard tasks used to test workers need to be changed frequently, since workers are prone to collude by discussing and learning the correct responses to the tasks and perform misleadingly well on tests while still making errors on actual work tasks.

There has been significant previous work on deriving worker ability estimates without Gold Standard tasks (see related work section). However, most of these papers do not provide confidence intervals for the estimates. Confidence intervals are essential to distinguish workers who are truly error prone from workers who were simply unlucky on a few tasks. For example, consider two scenarios. In the first scenario, a worker attempts 3 tasks and gets 1 of them wrong. In the second scenario, a worker attempts 30 tasks and gets 10 of them wrong. The best estimate for the worker error rate in both cases is $\frac{1}{3}$. Thus, existing methods that do not produce confidence intervals will judge the two workers as having the same ability. However, we can be much more sure that the worker in the second scenario has a high error rate. The first worker could simply have been unlucky with one of the tasks, and our error rate estimate of $\frac{1}{3}$ for the worker is less reliable. If we're going to fire a worker for having a high estimated error rate, then it is important to be sufficiently confident that the worker has low ability because firing many good workers can lead to a bad reputation in a crowdsourcing market, making it harder to attract workers in the future. Therefore, we focus on the problem of estimating worker abilities while providing confidence intervals for our estimates.

In a previous paper [2] we studied how to compute confidence intervals for worker error rates without a gold standard. The paper demonstrates that accurate confidence intervals can lead to significantly higher quality crowdsourced results, and can also decrease the time needed to find a pool of high quality workers. However, the paper makes some strong assumptions that limit its applicability. In particular, it assumes that *every worker attempts every task*—that is, if we are evaluating 100 workers across 100 (unsolved) tasks, then it assumes that we have a complete 100 × 100 matrix, which is unrealistic for two reasons: first, in many crowdsourcing platforms like Mechanical Turk [3], we cannot control which tasks a worker performs, nor can we force a worker to attempt every task, so it is unlikely that we will end up with a complete answer matrix. Second, it is unlikely that we would be willing to invest so much money in simply evaluating workers over using that money in solving more tasks. Additionally, [2] makes the assumptions that *tasks are binary* and that workers are *equally likely to make false positive and false negative errors*. Under these assumptions, the paper provides a mechanism to find confidence intervals for worker error rates and to leverage these intervals when hiring and firing workers. Reference [2] does provide a good starting point for computing confidence intervals, but due to the assumptions it makes, it ultimately ends up not being powerful enough for many real applications.

Our key contribution in this paper, in comparison with [2], is to produce *better (i.e., tighter) confidence intervals in a much more general setting*. Our confidence intervals are *better or tighter* in the sense that, unlike [2], where the confidence intervals may sometimes be excessively large (i.e. overly conservative), our confidence intervals match the accuracy values almost exactly. Our confidence intervals are *more general* in the sense that, unlike [2], where it is assumed that every worker attempts every task and that they are equally likely to make false positive and false negative errors, and every task has a boolean response, we make none of these assumptions. Since we are producing *better confidence intervals on a much more challenging setting*, straightforward adaptations of [2] do not apply, and we need entirely new techniques.

The way we compute confidence intervals is by comparing worker responses with those of other workers. In particular, if a worker disagrees with what other workers say on a particular task, then it may be a sign that the worker made a mistake. As we will see, to estimate confidence intervals using this intuitive idea, we must simultaneously estimate the error rates (and confidence intervals) of multiple workers, not all of whom have attempted every task. That is, we need to estimate multiple unknowns that interact in complex ways, which makes standard statistical techniques inapplicable to this problem, especially when we want to find confidence intervals that are "as tight as possible." Further, the number of variables to be estimated is even higher when tasks are $k$-ary instead of binary, as the number of probabilities to be estimated grow as $O(k^2)$.

One might think that we can deal with non-regular data (when not every worker has attempted every task) by restricting our attention to tasks that have been attempted by all workers, giving us a regular subset of the non-regular data. But the set of such tasks may be very small, especially on sparse data, and involves throwing away a large chunk of useful data, which is an inefficient use of the data available. Our techniques will make use of all the data available, while producing reliable confidence intervals for worker error rates.

The contributions of our paper are summarized below: First, we focus on providing better confidence intervals when not all workers have attempted each task.

A) We provide a technique for estimating worker ability with tight confidence intervals when there are multiple workers and data is non-regular (Section III-A, III-B, III-C).
B) We present experimental results on synthetic data (Section III-A, III-D) and real datasets (Section III-E) to study our confidence intervals when we have multiple workers and not every worker attempts every task.

Then, we focus on providing better confidence intervals when tasks are $k$-ary.

A) We provide a technique for estimating worker response probabilities with confidence intervals when tasks are $k$-ary (Section IV-A).
B) We present experimental results on synthetic data (Section IV-B) and 3 real datasets (Section IV-C) to study the performance of our confidence intervals on $k$-ary tasks.

## II. RELATED WORK

The prior work related to ours can be placed in a few categories; we describe each of them in turn:

**Crowd Algorithms:** There has been a lot of recent activity centered around designing data processing algorithms where the unit operations are performed by human workers, such as filtering [4], sorting and joins [5], [6], deduplication and clustering [7], [8], [9] and categorization [10], [11]. Evaluating worker ability accurately and then selecting workers based on ability before any of these algorithms are executed will surely improve eventual performance.

**Statistics:** In our work, we use many fundamental concepts from statistics [1], [12]. However, as we indicated in the introduction, standard techniques cannot be applied directly because we do not have ground truth answers and because we are simultaneously generating confidence intervals for multiple estimates (in this case, the worker error probabilities for each of the workers working on the same task) that are dependent on each other in complex ways.

**Expectation-Maximization:** Expectation Maximization, or EM [13], [14], [15], [16], [17] is a technique for finding maximum likelihood estimates for hidden parameters in statistical models. Expectation Maximization has been used to provide a maximum likelihood estimate for worker error rates by performing repeated iterations that converge to a locally optimal estimate. However, there is no guarantee that the convergence will be to actual worker error rates, that is, the globally optimal solution. Moreover, EM does not provide confidence intervals for the estimates generated.

**Heuristic Worker Error Estimation:** There has been significant work on simultaneous estimation of answers to tasks and errors of workers using the EM algorithm or other local optimization techniques. There have been a number of papers studying increasingly expressive models for this problem, including difficulty of tasks and worker expertise [18], [19], adversarial behavior [20], and online evaluation of workers [21], [22], [23]. These works provide point estimates and don't provide confidence guarantees for their estimates. Bayesian techniques have been applied to the problem of worker evaluation [24], but no theoretical guarantees on the accuracy of the evaluation are provided. There has also been work on choosing workers for evaluating different items so as to reduce overall error rate [25], [26]. While [26] uses heuristic confidence intervals in its algorithm, it does not provide confidence guarantees. The other papers do not provide confidence intervals of any kind.

**Optimal Worker Error Estimation:** Some work has been done on using algebraic techniques to evaluate workers or tasks, with theoretical guarantees. Reference [27] provides worker error rate estimates with upper bound guarantees on total error over all workers, but does not provide confidence intervals or any guarantee for individual workers. It also assumes that tasks have a binary output and that false positive and false negative error rates are equal (we make neither of these assumptions). There has also been work on finding correct labels for items in crowdsourcing, with theoretical guarantees [28], [29], but these works evaluate items rather than workers. Reference [28] also assumes that tasks are assigned to workers in a uniformly random fashion. Reference [2] is the only paper to provide confidence intervals for worker error rates, but assumes a very simple worker model, with binary tasks, regularity, and equal false positive and false negative error rates. On the other hand, we not only assume a highly expressive error model and task model, but we also provide confidence interval guarantees along with error rates, allowing users of our technique to have more fine-grained information to evaluate workers.

## III. GENERALIZING TO MULTIPLE WORKERS AND NON-REGULAR DATA

We present our techniques incrementally, starting with the simplest case of 3 workers with regular data.

### A. 3-worker Binary Regular

We have three workers $w_1$, $w_2$, $w_3$, and for each task $T_j$ ($1 \leq j \leq n$), we let $r_{1,j}$, $r_{2,j}$, $r_{3,j}$ denote the responses of $w_1$, $w_2$ and $w_3$ for task $j$. We assume that each response $r_{i,j}$ is binary Yes(Y)/No(N). Furthermore, for each task $T_j$, we assume that there is a true (correct) binary response $true(T_j)$. Note that $true(T_j)$ is not available when we are computing confidence intervals. Each worker $w_i$ has an inherent error rate $p_i$ representing the probability that the worker makes a mistake on a task. That is, $p_i = \Pr[r_{i,j} \neq true(T_j)]$. Thus, the probability that the worker makes a mistake on a task is assumed to be independent of other workers' responses for that task and for other tasks. This assumption is true as long as workers don't collude with each other, and tasks have equal difficulty. In reality, tasks do vary in difficulty, which creates a small amount of correlation between worker error probabilities; in Sections III-E and IV-C, we study real scenarios where independence does not hold, and we show that our methods are still very useful. In addition, we assume that workers are not malicious, i.e. $p_i < \frac{1}{2}$.

Our goal is to *compute confidence interval estimates $\hat{p}_1 \pm \epsilon_1$, $\hat{p}_2 \pm \epsilon_2$, $\hat{p}_3 \pm \epsilon_3$ for the worker error rates, given worker responses $r_{i,j}$*.

**Solution:** Our solution is similar to the method from [2], but we produce much tighter (less conservative) confidence intervals. As we will see in the next section, the confidence interval size is reduced by up to 40%.

The intuition is that even though we do not have the true responses for tasks, the rate of disagreements between workers across tasks is a good indicator of their error rate: a worker with a higher error rate is more likely to disagree with others. For instance, two workers with a zero error rate will never disagree, while a worker with an error rate close to $\frac{1}{2}$ will disagree with other good workers about half the time. Thus, we find confidence intervals by first computing pairwise disagreements and translating them into error rate estimates.

We define fraction of agreements for each pair of workers to be $\hat{q}_{i,j}$, while $Q_{i,j}$ is defined to be the corresponding random variable. Let $q_{i,j}$ denote the expected agreement rate $E[Q_{i,j}]$. For example, if the responses of workers $w_1$ and $w_2$ agree for 40 tasks but disagree in 10 tasks, then $\hat{q}_{1,2}$ is $\frac{40}{50}$. The expected agreement rates $q_{i,j}$ are related to the worker error rates $p_i$ as follows: $q_{i,j} = p_i p_j + (1-p_i)(1-p_j)$ for $(i,j) = (1,2),(2,3),(1,3)$. This is because two workers agree on a task if and only if they are both right or both wrong on the task. Solving these equations for $p_1, p_2, p_3$ gives us:

$$p_1 = \frac{1}{2} - \frac{1}{2}\sqrt{\frac{(2q_{1,2}-1)(2q_{1,3}-1)}{2q_{2,3}-1}} \quad (1)$$

with analogous equations for $p_2$ and $p_3$. We can use these equations to translate our estimates for $q_{i,j}$ (i.e. $\hat{q}_{i,j}$) into error rate estimates. However, we need confidence intervals for the error rates rather than point estimates.

To get confidence intervals, we use the following theorem. (In Theorem 1 as well as the rest of the paper, we use Cov to denote Covariance and Dev to denote Standard Deviation [1].)

**Theorem 1.** Suppose we have $k$ approximately normal random variables $X_1, \ldots, X_k$, such that $\forall i, 1 \leq i \leq k : e_i = E[X_i]$ and $\forall i,j, 1 \leq i,j \leq k : c_{i,j} = \text{Cov}(X_i, X_j) = E[X_i X_j] - E[X_i]E[X_j]$. Also suppose we have a locally linear function $f$, such that $f(e_1 + a_1, \ldots, e_k + a_k) \approx f(e_1, \ldots, e_k) + \sum_{i=1}^{k} d_i a_i$, for some constants $d_1, \ldots, d_k$. Then, for the random variable $Y = f(X_1, \ldots, X_k)$, we have

$$E[Y] \approx f(e_1, \ldots, e_k) \qquad \text{Dev}(Y) \approx \sqrt{\sum_{i=1}^{k}\sum_{j=1}^{k} d_i d_j c_{i,j}}$$

and the $c$-confidence interval of $Y$ is given by:

$$CI(Y,c) \approx (E[Y] - z_t \text{Dev}(Y), E[Y] + z_t \text{Dev}(Y)) \quad (2)$$

where $z_t$ is the $t^{th}$ percentile of the normal distribution, for $t = \frac{1-c}{2}$, and $CI$ stands for Confidence Interval. $\square$

For the 3-worker binary case, we will use Theorem 1 once for each $p_i$ where (a) the $X_i$ are the $Q_{i,j}$, (b) $Y$ is the $p_i$ and (c) the function $f$ corresponds to the equation

$$f(\hat{q}_{1,2}, \hat{q}_{3,1}, \hat{q}_{2,3}) = \frac{1}{2} - \frac{1}{2}\sqrt{\frac{(2\hat{q}_{1,2}-1)(2\hat{q}_{3,1}-1)}{2\hat{q}_{2,3}-1}}$$

Since the $Q_{i,j}$s are a sum of iid Bernoulli random variables, we can assume they are approximately normal, which is a requirement of Theorem 1.

Thus, to apply theorem 1 to our setting, we need estimates (a) $\hat{q}_{i,j}$ of the $Q_{i,j}$s, (b) covariances between $Q_{i,j}$s, and (c) the partial derivatives of the function $f$ with respect to each of its arguments. Estimates for (a) are readily available.

*Requirement (b) (covariances):* Each $Q_{i,j}$ is the sum of iid Bernoulli random variables, with one variable per task, taking value 1 if workers $w_i$ and $w_j$ give the same response on the task and 0 otherwise. $Q_{i,j}$ and $Q_{i,k}$ are expected to be positively correlated, because an agreement between workers $w_i$ and $w_j$ on a task makes it more likely that they are both correct, increasing the chance of workers $w_i$ and $w_k$ agreeing on the task. These covariances are estimated using Lemma 1.

**Lemma 1.** The Covariances of the $Q_{i,j}$s are obtained as follows: For $i,j \in \{1,2,3\}, i \neq j : \text{Cov}(Q_{i,j}, Q_{i,j}) = \frac{q_{i,j}(1-q_{i,j})}{n}$. And for distinct $i,j,k \in \{1,2,3\}$,

$$\text{Cov}(Q_{i,j}, Q_{j,k}) = \frac{p_j(1-p_j)(2q_{i,k}-1)}{n} \quad \square$$

While we do not have the actual values $q_{i,j}$ and $p_i$ available to us, we can obtain approximate values of the covariances using estimates $\hat{q}_{i,j}$ and $\hat{p}_i$.

*Requirement (c) (partial derivatives):* We now focus on (c). Our function is partially differentiable with respect to each of its arguments and thus is locally linear, with the partial derivatives as linear coefficients.

$$f(e_1 + a_1, e_2 + a_2, \ldots) \approx f(e_1, e_2, \ldots) + \sum_i a_i \frac{\partial f(e_1, e_2, \ldots)}{\partial e_i}$$

We compute these coefficients using Lemma 2.

**Lemma 2.** Let $f(q_{i,j}, q_{i,k}, q_{j,k}) = \frac{1}{2} - \frac{1}{2}\sqrt{\frac{(2q_{i,j}-1)(2q_{i,k}-1)}{2q_{j,k}-1}}$. Then, the partial derivatives of $f$ with respect to its inputs are given by:

$$\frac{\partial f(q_{i,j}, q_{i,k}, q_{j,k})}{\partial q_{i,j}} = -\sqrt{\frac{(q_{i,k} - \frac{1}{2})}{8(q_{i,j} - \frac{1}{2})(q_{j,k} - \frac{1}{2})}}$$

$$\frac{\partial f(q_{i,j}, q_{i,k}, q_{j,k})}{\partial q_{i,k}} = -\sqrt{\frac{(q_{i,j} - \frac{1}{2})}{8(q_{i,k} - \frac{1}{2})(q_{j,k} - \frac{1}{2})}}$$

$$\frac{\partial f(q_{i,j}, q_{i,k}, q_{j,k})}{\partial q_{j,k}} = \sqrt{\frac{(q_{i,j} - \frac{1}{2})(q_{i,k} - \frac{1}{2})}{8(q_{j,k} - \frac{1}{2})^3}} \quad \square$$

*Complete Method:* Our method for computing confidence intervals is shown in Algorithm A1. In Step 1, 2, and 3, we compute the requirements as described above. Finally, in Step 4, we solve for the $\hat{p}$ values and compute the confidence intervals using Theorem 1 and the covariances and derivatives from Steps 2 and 3.

---

**Algorithm A1:**

1) For each pair of workers $w_i$ and $w_j$, compute the fraction of agreements $q_{i,j}$
2) Compute covariances for pairs of $Q_{i,j}$ using Lemma 1
3) Compute the partial derivative of $f$ with respect to each $q_{i,j}$ using Lemma 2.
4) With $p_1 = f(q_{1,2}, q_{1,3}, q_{2,3})$, $p_2 = f(q_{1,2}, q_{2,3}, q_{1,3})$, $p_3 = f(q_{1,3}, q_{2,3}, q_{1,2})$, and the derivatives and covariances computed in steps 2 and 3, apply Theorem 1 to get confidence intervals for $p_1$, $p_2$ and $p_3$.

---

*1) Comparison with the Old Technique:* Before moving on to generalizations of our technique, we present an experiment to compare the compactness of our confidence intervals to the confidence intervals from [2]. We use data from simulations for this experiment. (Experiments on real crowds is presented later on.) We fix the number of tasks $n = 100$ (we observe the relative performance of the two techniques does not depend on $n$), the number of workers $m$, and confidence level $c$. We have a set of tasks $T_1, \ldots, T_n$, and a set of workers $w_1, \ldots, w_m$. The error rate $p_i$ of each worker $w_i$ is independently chosen to be one of 0.1, 0.2 or 0.3 with equal probability. Whenever worker $w_i$ attempts a task, he or she makes a mistake with the chosen probability $p_i$.

We refer to the technique in this paper as the 'new technique', while the technique in [2] is the 'old technique'. After generating simulated data, we apply both the new technique and the old technique to generate two sets of $c$-confidence intervals, for each worker's error rate. We do this 500 times and compute the average interval size for each technique. This average size is computed for both the old and new techniques, for every value of $c \in \{0.05, 0.1 \ldots 0.95\}$, and is plotted against $c$ in Figure 1. We depict the trend for $m \in \{3, 7\}$.

The plot shows that the interval sizes produced by the new technique are significantly smaller for equal values of $n$, $m$ and $c$. For instance, for $n = 100$, $m = 3$, $c = 0.5$, the average interval size for the old technique is around $0.11$, while the average interval size for the new technique is $0.07$, giving almost a 40% size reduction—representing a much more precise understanding of worker quality.

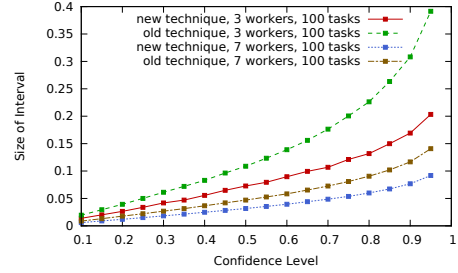

Fig. 1. Size of interval vs. confidence for old and new techniques

Moreover, the accuracy of these tighter confidence intervals is very high, as we will see subsequently. (An $X$ % confidence interval is accurate if when we repeat the experiment 100 times, in $X$ or more cases the true mean is in the interval.) We do not show accuracy results here since they are very similar to those that will be shown in Figure 2 for a more general case. Thus, we do not lose anything by using the 'new technique'; in fact, we only tighten our worker error rate confidence intervals significantly, which can be tremendously useful.

### B. 3-worker Binary Non-Regular

We now move to the 3-worker case where each worker has done some, but not necessarily all of the tasks. As such, we have three workers $w_1, w_2, w_3$, and for each task $T_j$, there are three responses $r_{1,j}, r_{2,j}, r_{3,j}$ (If worker $w_i$ has not attempted task $T_j$, then $r_{i,j}$ is undefined). We still assume that each pair of workers has done at least one task in common, as otherwise we do not have enough data to compute the error rates. The number of tasks attempted by both workers $w_i$ and $w_j$ is denoted by $c_{i,j}$, and the number of tasks attempted by all three workers is denoted by $c_{i,j,k}$. Thus, if there are 100 tasks, of which $w_1$ has attempted the first 80, $w_2$ has attempted the last 80, and $w_3$ has attempted the middle 80, then $c_{1,2} = 60$, $c_{1,3} = c_{2,3} = 70$, $c_{1,2,3} = 60$.

**Solution:** To compute confidence intervals, we need the fraction of agreements $q_{i,j}$ for each pair of workers, as we did in Section III-A. But this time, the fraction is only taken over the tasks attempted by both workers in the pair. For example, if workers $w_1$ and $w_2$ have attempted 60 tasks in common (some of which may have been attempted by $w_3$ as well), and their responses agree on 50 of the tasks, then $q_{1,2} = \frac{50}{60}$.

Our method for computing confidence intervals is same as the one in Algorithm A1, except that the covariances between the $Q_{i,j}$s now depend on how many tasks workers have done in common with each other. For example, if workers $w_1$ and $w_3$ have many tasks in common with $w_2$, but very few in common with each other, then we would expect $Q_{1,2}$ and $Q_{2,3}$ to be very weakly correlated, since they are computed over nearly disjoint sets of tasks. We use Lemma 3 instead of Lemma 1, to compute covariances in Step 2 of Algorithm A1. (Note that Lemma 1 is actually a special case of Lemma 3, with $c_{1,2} = c_{2,3} = c_{1,3} = n$, where $n$ is the number of tasks. )

**Lemma 3.** The Covariances of the $Q_{i,j}$s are obtained as follows: For $i,j \in \{1,2,3\}, i \neq j$,

$$\mathrm{Cov}(Q_{i,j}, Q_{i,j}) = \frac{q_{i,j}(1 - q_{i,j})}{c_{i,j}}$$

And for distinct $i, j, k \in \{1, 2, 3\}$,

$$\text{Cov}(Q_{i,j}, Q_{j,k}) = \frac{c_{i,j,k} \cdot p_j(1-p_j)(2q_{i,k}-1)}{c_{i,j}c_{j,k}} \qquad \square$$

The computational complexity of our method for evaluating 3 workers is $O(n)$, where $n$ is the number of tasks.

### C. m-worker Binary Non-Regular

Next we consider the case with $m (\geq 3)$ workers, where each worker may not have attempted every task. As such, we have $m$ workers $w_1, w_2, \ldots w_m$, and for each task $T_j$ there are $m$ responses $r_{1,j}, r_{2,j}, \ldots r_{m,j}$ (As before, if worker $w_i$ has not attempted task $T_j$, then $r_{i,j}$ is undefined). The number of tasks attempted by both workers $w_i$ and $w_j$ is denoted by $c_{i,j}$, and the number of tasks attempted by each of three workers $w_i, w_j, w_k$ is $c_{i,j,k}$. In this case, our goal is to get confidence intervals $\hat{p_1} \pm \epsilon_1, \hat{p_2} \pm \epsilon_2 \ldots \hat{p_m} \pm \epsilon_m$ for the worker error rates.

**Solution:** To evaluate a worker $w$, we will take two other workers and apply our 3-worker method from Section III-B to the resulting 3 workers to get an error rate estimate for worker $w$. We repeat this process several times, each time combining worker $w$ with two different workers and getting additional estimates for the error rate of worker $w$. We will then combine all these estimates and use Theorem 1 to get a confidence interval for worker $w$'s error rate.

We now describe the method (listed in Algorithm A2) in detail below. The method runs for $m$ iterations, computing the confidence interval for one worker per iteration.

*Step 1: Form pairs:* Suppose the worker we're trying to evaluate is $w_i$. We start by dividing the remaining workers into pairs (which along with worker $w_i$ will form a triple). Although pairing will give us a correct answer, the way we form the pairs will impact the size of the final confidence interval. We discuss the strategy for forming pairs in Section III-C1.

*Step 2: Apply 3-worker method per triple:* Suppose we formed $l$ pairs. To each pair, add the worker $w_i$ to form a triple. Let the $k^{th}$ triple be denoted by $Triple_k$. Let the workers in $Triple_k$ be $w_i, w_{j_1}$, and $w_{j_2}$.

For each triple, we apply a portion of our 3-worker method from Section III-B (In fact, all but computing the confidence interval in Theorem 1 is performed.):

- We first compute the agreement rates $q_{i,j_1}, q_{i,j_2}$ and $q_{j_1,j_2}$ as we did in the 3-worker case. Let $f(a,b,c) = \frac{1}{2} - \frac{1}{2}\sqrt{\frac{(2a-1)(2b-1)}{2c-1}}$ as before. The agreement rates allow us to compute the estimated error rate $p_{k,i}$ as $f(q_{i,j_1}, q_{i,j_2}, q_{j_1,j_2})$.
- We then compute the partial derivative of $f(q_{i,j_1}, q_{i,j_2}, q_{j_1,j_2})$ with respect to $q_{i,j_1}$ (call the derivative $d_{k,i,j_1}$) and $q_{i,j_2}$ (call the derivative $d_{k,i,j_2}$) using Lemma 2. These derivatives and the covariances allow us to get the standard deviation for the estimate of $p_{k,i}$, which we call $\text{Dev}_{k,i}$.

Thus, at the end of the procedure, we have estimates for error rates $p_{k,i}$, partial derivatives $d_{k,i,j_2}$, as well as deviation estimates for the error rate estimate $\text{Dev}_{k,i}$.

Now, instead of directly completing the application of Theorem 1 using Equation 2 to give a confidence interval using information from just one triple, we instead aggregate the estimates from various triples, giving much tighter estimates.

*Step 3: Aggregating information from triples:* We now turn once again to Theorem 1 to aggregate information from various triples. We let our new function $f'(p_{1,i}, \ldots, p_{l,i}) = \sum_{k=1}^{l} a_k p_{k,i}$, where $a_k$ are some weights on the $p_{k,i}$. We describe how to select $a_k$ later on; a uniformly weighted average would suffice as well, but the size of the final confidence interval can be reduced significantly by optimizing these weights, as we will describe in Section III-C1.

For now, in order to apply Theorem 1 on $f'$, we already have error rate estimates $p_{k,i}$, partial derivatives $d_{k,i,j_2}$, as well as deviation estimates for the error rate estimate $\text{Dev}_{k,i}$. The only missing ingredient is the covariances, $\text{Cov}_{k_1,k_2}$ between each pair of estimates $p_{k_1,i}$ and $p_{k_2,i}$, which we describe how to compute next, using Lemma 4.

**Lemma 4.** Let the error rate estimate for worker $w_i$ in its $k^{th}$ triple, be denoted by $p_{k,i}$, and its deviation be $\text{Dev}_{k,i}$. For $k_1, k_2$ such that $1 \leq k_1 \leq k_2 \leq l$, suppose $Triple_{k_1}$ contains workers $w_i, w_{j_1}, w_{j_2}$, and $Triple_{k_2}$ contains workers $w_i, w_{j_3}$ and $w_{j_4}$. For all pairs of workers $w_j, w_{j'}$, let $q_{j,j'}$ denote their agreement rate, and let $c_{j,j'}$ denote the number of tasks they've attempted in common, and for all triples of workers $w_j, w_{j'}, w_{j''}$, let $c_{j,j',j''}$ be the number of tasks attempted by all three workers in common. Let $d_{k,i,j}$ denote the partial derivative of $p_{k,i}$ with respect to $q_{i,j}$. Then the Covariances of the $p_{k,i}$s are obtained as follows: For $k_1 = k_2$:

$$\text{Cov}(p_{k_1,i}, p_{k_2,i}) = \text{Dev}_{k_1,i}\text{Dev}_{k_1,i}$$

And for $k_1 \neq k_2$, $\text{Cov}(p_{k_1,i}, p_{k_2,i})$ is equal to

$$d_{k_1,i,j_1}d_{k_2,i,j_3}C(i,j_1,j_3) + d_{k_1,i,j_1}d_{k_2,i,j_4}C(i,j_1,j_4) + \\ d_{k_1,i,j_2}d_{k_2,i,j_3}C(i,j_2,j_3) + d_{k_1,i,j_2}d_{k_2,i,j_4}C(i,j_2,j_4)$$

where $C(i,j,j')$ is defined as

$$C(i,j,j') = \frac{c_{i,j,j'}p_i(1-p_i)(2q_{j,j'}-1)}{c_{i,j}c_{i,j'}} \qquad \square$$

If the total number of tasks is $n$, and number of workers being evaluated simultaneously is $m$, the computational complexity of our algorithm for evaluating $m$ workers is $O(m^2 n + m^4)$. This can be reduced to $O(m^2 n + m^{3.373})$ if matrix inversion is done using William's Algorithm instead of Gauss-Jordan elimination. Since we take a square root during the confidence interval computation, there is a minuscule probability that our algorithm fails due to a negative value occurring under the square root. For instance, a worker may have error rate $p_i < \frac{1}{2}$ but there is still a non-zero probability of the worker getting all responses wrong. However, the probability of our algorithm failing falls exponentially with the number of tasks.

---

**Algorithm A2:**

For each worker $w_i$ do:

1) Divide the remaining workers into pairs, with one worker possibly left over. Each pair along with $w_i$ forms a triple $Triple_k$. Let the number of triples be $l$.
2) For each triple $Triple_k$, let the members of the triple be $w_i, w_{j_1}$ and $w_{j_2}$. Then for $Triple_k$, do:
   a) Compute the agreement rates $q_{i,j_1}, q_{i,j_2}$ and $q_{j_1,j_2}$

---

b) Compute the partial derivative of $f(q_{i,j_1}, q_{i,j_2}, q_{j_1,j_2})$ with respect to each of $q_{i,j_1}$, $q_{i,j_2}$ and $q_{j_1,j_2}$ using Lemma 2. Let $d_{k,i,j_1}$ be the derivative with respect to $q_{i,j_1}$, $d_{k,i,j_2}$ be the derivative with respect to $q_{i,j_2}$ and $d_{k,j_1,j_2}$ be the derivative with respect to $q_{j_1,j_2}$

c) Using the 3-worker procedure from Section III-B, compute the confidence interval for worker $w_i$. Let the mean of the interval be $p_{k,i}$ and compute the deviation $\text{Dev}_{k,i}$ as returned by Theorem 1 in the method

For $\text{Triple}_k$, we now have an estimate $p_{k,i}$, the standard deviation for the estimate $\text{Dev}_{k,i}$ and partial derivatives $d_{k,i,j_1}, d_{k,i,j_2}, d_{k,j_1,j_2}$

3) For each pair of triples $\text{Triple}_{k_1}, \text{Triple}_{k_2}$: find the Covariance $\text{Cov}_{k_1,k_2}$ between the estimated error rates $p_{k_1,i}$ and $p_{k_2,i}$, using Lemma 4. Then, use it, along with previously computed quantities to estimate worker $w_i$ error rate $p_i$ and confidence interval using Theorem 1 on $p_i = f'(p_{1,i}, p_{2,i}...p_{l,i}) = \sum_{k=1}^{l} a_k p_{k,i}$.

In [2], the worker evaluation technique essentially divided the other workers (apart from the one under consideration) into two disjoint sets of workers, then treated as super-workers. That is, the response of a super-worker equals the majority response of the workers in that set. Unfortunately, the super-worker technique does not apply in the setting where not every worker has attempted every task, since the technique relies on a fundamental assumption that the super-worker must have a consistent error rate across all tasks. Here, in a non-regular setting, since different subsets of workers attempt different tasks, this assumption is violated.

*1) Optimizations for the $m$-worker method :* The method given in the previous section produces correct confidence intervals. But the size of the confidence intervals can be improved by optimizing two steps of the method, namely splitting workers into pairs (Step 1) and choosing linear weights for combining the estimates from different triples (Step 3).

*Selecting triples:* The quality of an estimate from a triple is higher when the workers in the triple have attempted more tasks in common. Moreover, because we can give different weight to the estimate from different triples, it is better to have some very good triples, and other bad triples, than to have many average triples. So we construct the pairs of workers using a greedy approach. Suppose $w_i$ is the worker to be evaluated. We create a list of all workers other than $w_i$ that we sorted in descending order of the number of tasks they have attempted in common with $w_i$. Then we take the first member of the list, say $w_{i_2}$, and pair him with the first worker in the rest of the list to have at least one common task with both $w_i$ and $w_{i_2}$, say $w_{i_3}$. This gives us a pair $(w_{i_2}, w_{i_3})$. The members of the pair are then removed from the list. We then repeat the process on the remaining list to find the next pair, until the list has no more pairs of workers who have a common task with $w_i$ and with each other.

*Setting $a_k$:* We now describe how to set the weights in the function $f'$. There are $l$ triples $\text{Triple}_1, \text{Triple}_2...\text{Triple}_l$ in totla, giving $l$ estimates $p_{1,i}, p_{2,i}, ...p_{l,i}$. For each pair of triples $\text{Triple}_{k_1}, \text{Triple}_{k_2}$, we have the covariance of their estimates $\text{Cov}_{k_1,k_2}$. Let $C$ be the covariance matrix i.e., $C(k_1, k_2) = \text{Cov}_{k_1,k_2}$. Suppose we take our final estimate to be: $\sum_{k=1}^{l} a_k p_{k,i}$, where $\sum_{k=1}^{l} a_k = 1$. Thus the $a_k$s are our linear weights. Let $A$ be the $l \times 1$ weight matrix, $A(k, 1) = a_k$. Then, the variance of the final estimate is given by: $A^T C A$.

The size of the final confidence interval will be proportional to the square root of the variance, so our aim is to choose linear weights $a_k$s to minimize the variance. Lemma 5 shows how the variance can be minimized.

**Lemma 5.** Given an $l \times l$ matrix $C$, the $l \times 1$ matrix $A$ such that $\sum_{k=1}^{l} A(k) = 1$, which minimizes $A^T C A$ is obtained as follows: Let $O$ be the $l \times 1$ matrix with all entries equal to 1 and let $B = C^{-1}O$. Then the optimum value of $A$ is given by $A = \frac{B}{\|B\|_1}$, where $\|B\|_1$ is the $L_1$-norm of $B$. □

Using the $A$ obtained from the lemma, we get the optimal linear weights with which to combine the results from the different triples.

### D. Experiments on Synthetic Data

We now move on to an experimental evaluation of our confidence intervals. We present two types of experiments: experiments on synthetic (simulated) data and those on real world data. The experiments on synthetic data, presented in this section, allow us to study the performance of our intervals on a wide variety of parameters, while real world data experiments (presented in the section III-E) allow us to test the usefulness of our techniques in practice.

We first experimentally evaluate our confidence intervals using data from simulations. In each experiment, we have a set of tasks $T_1, \ldots, T_n$, and a set of workers $w_1, \ldots, w_m$. The error rate $p_i$ of each worker $w_i$ is chosen to be one of $0.1$, $0.2$ or $0.3$ with equal probability, independently of error rates of other workers. Whenever worker $w_i$ attempts a task, he or she makes a mistake with probability $p_i$, independently of any other attempts on any task.

*1) Confidence vs. Accuracy:* We first fix the number of tasks $n$, the number of workers $m$, and a confidence level $c$. For every worker-task pair, the worker attempts that task with probability $0.8$, independently of which other tasks the worker has attempted or which tasks other workers have attempted (We tried many values for the probability with which a worker attempts a task and obtained similar results). This gives us our non-regular data. We run our $m$-worker binary non-regular scheme on this data to get $c$-confidence intervals for the worker error rates. For each interval computed, we check if the true error rate of the worker being evaluated lies within the confidence interval. We repeat the experiment 500 times, with different sets of workers, and compute the interval-accuracy i.e., the number of intervals that contained the true worker error rate divided by the total number of intervals. This interval-accuracy is computed for every value of $c \in \{0.05, 0.1, 0.15...0.95\}$ and plotted against $c$ in Figure 2(a). We generate a plot for $n \in \{100, 300\}$ and $m \in \{3, 7\}$.

Note that the solid line $y = x$ represents the ideal accuracy level. If the accuracy is above the line, then it means that our intervals are larger than they need to be, whereas if it is below the line, it means that our intervals are not large enough. The plot shows that our method does almost achieve the ideal level of accuracy. Our method provably achieves ideal accuracy if the two assumptions in Theorem 1 (input random variables are normally distributed, function is locally linear) hold. These assumptions do not hold exactly even in the simulated data,

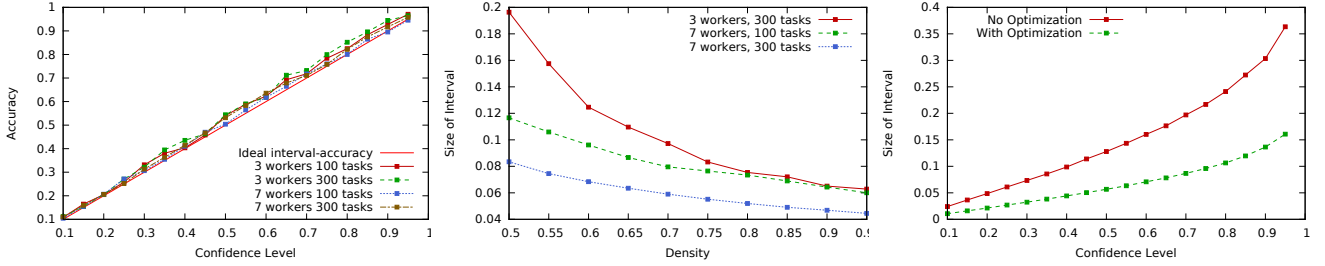

Fig. 2. (a)Accuracy of $m$-worker binary non-regular method in estimating confidence (b) Size of intervals for varying levels of density (c)Size of interval vs. confidence with and without weight optimization

because the agreement rates $Q_{i,j}$ are binomially distributed rather than normally, and the function $f$ we use is not linear. But the experiment shows that our method still almost achieves the ideal accuracy level.

*2) Density vs. Interval Size:* We now study the relation between the density of data (the fraction of worker-task pairs in which the worker actually attempted the task) and the confidence interval size. We fix the number of tasks $n$, the number of workers $m$, and a density $d$. We also fix confidence level $c$ to be $0.8$. Each worker attempts each task with probability $d$, independently of which other tasks the worker has attempted, and which tasks other workers have attempted. We run our $m$-worker binary non-regular method on the resulting data to obtain $c$-confidence intervals for the error rate of each worker. We repeat this experiment 500 times and compute the average size of all the confidence intervals generated. This average size is computed for every value of $d \in \{0.5, 0.55, 0.6...0.95\}$ and plotted against $d$ in Figure 2(b). We make such a plot for $(n,m) \in \{(100,7), (300,3), (300,7)\}$. There is no plot for $(100,3)$ because the sizes are too large to fit with the current scale at $d = 0.5$.

As density increases for fixed values of $m$, $n$, and $c$, the average size of our $c$-confidence intervals is expected to decrease since more density means we have more data available for our estimate. Figure 2(b) confirms this. Moreover, interval size seems to be inversely proportional to density. This can be explained as follows: The standard deviation of our estimate of a Bernoulli random variable is inversely proportional to the number of instances of the variable available to us. In our method, we start by estimating the Bernoulli random variables $Q_{i,j}$ which give the agreement probability of a pair of workers and use those estimates to get confidence intervals for worker error rates. The number of instances of the $Q_{i,j}$ variables available is equal to the number of worker-task pairs, which is proportional to the square of the density. Hence the deviation of our estimate and the average size of our confidence interval is inversely proportional to density.

*3) Significance of Weight Optimization:* In Section III-C1, we saw how the linear weights for combining estimates from different triples can be optimized. Setting all the weights to $\frac{1}{l}$, where $l$ is the number of triples, gives valid confidence intervals (a $c$-confidence interval for a worker error rate is valid if it contains the true error rate with probability at least $c$), but the interval size for such weights can be larger than the size obtained on using the optimal weights.

In this experiment, we compare the performance of our confidence intervals using the optimal weights with their performance when using uniform weights. We fix the number of tasks $n$ to be 100 because we observe that changing the number of tasks does not change the relative size of the optimized and unoptimized intervals. We fix the number of workers $m$ to be 7. We do not consider 3 workers unlike in the other experiments, because with 3 workers, there would be only one triple, making any optimization meaningless. In addition, we set the density $d_i$, which is the fraction of tasks attempted by worker $w_i$, to $\frac{0.5i+(m-i)}{m}$. This ensures that different workers have attempted different numbers of tasks, and hence different triples will produce estimates of different quality, making the weight optimization necessary. (We also observe that in real world scenarios, the number of tasks attempted by different workers can vary widely). We then fix a confidence level $c$ and have each worker $w_i$ attempt each task with probability $d_i$, independently of which other tasks the worker has attempted or which tasks other workers have attempted. We then run our $m$-worker binary non-regular method on the resulting data to generate $c$-confidence intervals for the error rate of each worker. We repeat this experiment 500 times and compute the average size of all the confidence intervals generated. This average size is computed for every $c \in \{0.05, 0.1...0.95\}$ and plotted against $c$ in Figure 2(c).

The plot shows that interval sizes produced when using optimized weights are much smaller than the corresponding sizes are when using uniform weights. For example, the $50\%$ confidence interval using optimized weights has size about $0.05$, while the $50\%$ confidence interval obtained using uniform weights has size about $0.12$, which is more than twice the size using optimized weights. Thus it appears that optimizing the linear weights is crucial to getting compact confidence intervals.

*E. Experiments on Real Data*

We now evaluate our confidence intervals on real data to test their performance in settings where our assumptions such as uniform task difficulty, and non-collaboration of workers, may not hold. We use three different datasets, which we call Image Comparison (IC), Entailment (ENT), and Temporal (TEM). Gold standard responses are known for all the tasks in every dataset. Since we do not know the true error rate of the workers, we use the Gold Standard responses to compute the fraction of answers each worker got wrong and use that as a proxy for the worker's error rate. The datasets and experiments are described in the sections below.

*1) Data and Setting:* The first dataset, IC, is from [2]. In this dataset, a task consists of a pair of sports photos, containing one person each, and the worker is asked to state whether the photos contain the same person. There are 48 tasks, each of which was attempted by 19 workers on Amazon's Mechanical Turk [3]. Hence this dataset is regular, giving a total of $48 \times 19$ responses. To make the dataset non-regular, we randomly remove $20\%$ of the responses while performing experiments.

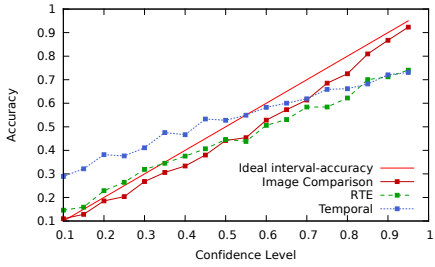

Fig. 3. Accuracy of interval vs confidence

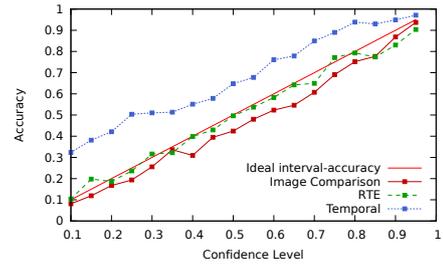

Fig. 4. Accuracy of improved interval vs confidence

The other two datasets, ENT and TEM, are from [30]. In ENT, a task consisted of a pair of sentences, and the worker had to state if the first sentence entailed the second. There were a total of 800 pairs of sentences and 164 different workers, but each worker did not attempt every task.

In TEM, a task consisted of a pair of sentences, and the worker had to identify if the event given in the first sentence temporally preceded the event in the second sentence. For instance, given the pair of sentences "Sam fell. Bob pushed him.", the true response would be 'No' because the event in the second sentence happened before the event in the first sentence. There were a total of 462 tasks and 76 workers, but not every worker attempted every task. The datasets ENT and TEM are not only non-regular, but are also rather sparse, with only a small fraction of workers attempting every task.

*2) Confidence vs. Accuracy:* We fix a confidence level $c$, and apply our $m$-worker binary non-regular method to get $c$-confidence intervals for error rates. For each interval, we check if the true error rate of the worker being evaluated lay within the confidence interval, and then compute the fraction of intervals which contained the true error rate. This fraction, called "Interval-Accuracy", is computed for every value of $c \in \{0.05, 0.1 ... 0.95\}$ and plotted against $c$ in Figure 3.

The solid line $y = x$ shows the ideal accuracy value, which is equal to the confidence level. The figure shows that the accuracy values of our intervals are reasonably close to the ideal values. Points above the $x = y$ line indicate that our resulting confidence intervals were conservative, i.e., slightly larger than necessary. More worrisome are points below the $x = y$ line where our interval was not large enough. This later divergence becomes greater when the desired confidence level is close to 1. We will next show how this can be remedied.

Our model had assumed that worker error rates are $< \frac{1}{2}$ (workers are not malicious), and hence agreement rates are also $> \frac{1}{2}$. However, this assumption is sometimes violated (possibly due to random fluctuations) in the data. The function $f(a, b, c) = \frac{1}{2} - \frac{1}{2}\sqrt{\frac{(2a-1)(2b-1)}{2c-1}}$ used in computing the worker error rates has a singularity at $c = \frac{1}{2}$ and is more volatile at values close to $\frac{1}{2}$. Thus our method performs badly when worker agreement rates are close to $\frac{1}{2}$. To remedy this, we preprocess the data and prune out workers who are almost surely spammers (and hence have error rates close to $\frac{1}{2}$). This can be done by running a simple majority technique, i.e., computing the fraction of times each worker disagrees with the majority, and using that as an approximation of the error rate. Once we have this approximate error rate, we remove workers whose approximate error rate is $> 0.4$ (since they are almost surely pure spammers). Then we run the $m$-worker binary non-regular method on the rest of the workers to get confidence intervals on their error rates as before and plot the interval accuracy vs confidence level $c$, in Figure 4. As the figure shows, removing the pure spammers leads to a significant improvement in the accuracy of our confidence intervals, especially at higher values of $c$.

## IV. GENERALIZING TO NON-BINARY TASKS

### A. *3-worker k-ary non-regular*

We now consider the case where tasks are $k$-ary instead of binary, for some integer $k$. Thus for every task, there are $k$ possible responses $r_1, r_2, ... r_k$. We have 3 workers $w_1$, $w_2$, and $w_3$, and each worker may not have attempted every task. Thus for every task $T_j$, we have up to 3 responses $r_{1,j}$, $r_{2,j}$, and $r_{3,j}$. Each $r_{i,j}$ is either undefined (if the worker did not attempt that task) or one of $r_1, r_2, .. r_k$. We sometimes also use the null response $r_0$ to denote that the worker did not attempt the task. In addition, each task $T_j$ has a true $k$-ary response True($T_j$) that is unknown to us.

We assume that each worker $w_i$ has a $k \times k$ "response probability" matrix $P_i$, where entry $P_i(j_1, j_2)$ in the $j_1^{th}$ row and $j_2^{th}$ column of $P_i$ gives the probability that worker $w_i$ responds with $r_{j_2}$ when the true answer is $r_{j_1}$. For instance, a worker may be biased towards giving response $r_1$, and thus their matrix will have higher values in the first column of the response probability matrix. From the definition of $P_i(j_1, j_2)$, we have $\sum_{j_2=1}^{k} P_i(j_1, j_2) = 1$. The probability that worker $w_i$ responds with $r_{j_2}$ to a task with true response $r_{j_1}$, is independent of $w_i$s responses to any other tasks or any other worker's responses to any task. We assume that $P_{j_1,j_1} > P_{j_1,j_2}$ for any $j_1 \neq j_2$. Our goal is to compute confidence intervals for $P_i(j_1, j_2)$ for each $1 \leq i \leq 3$, $1 \leq j_1, j_2 \leq k$.

In addition, let $S$ be the $k \times 1$ selectivity matrix. That is, the $i^{th}$ entry of $S$ gives the a priori probability that the true response to a task is $r_i$. Hence $\sum_{i=1}^{k} S(i) = 1$. Let $S_D$ denote the diagonal form of $S$, i.e. $S_D(i, j) = S(i)$ if $i = j$ and 0 otherwise. Also let $S_D^{\frac{1}{2}}$ denote the element-wise square root of $S_D$. We assume that $S$ is not known to our algorithm. However, our algorithm finds estimates for $S^{\frac{1}{2}}P_1$, $S^{\frac{1}{2}}P_2$ and $S^{\frac{1}{2}}P_3$. Then using the fact that the sum of values in each row of each $P_i$ is 1, we can estimate both the selectivity matrix $S$ and the response probability matrices $P_i$.

Our method for computing confidence intervals is shown in Algorithm A3. Before applying the method we pre-process the data, by building a 3-dimensional $(k+1) \times (k+1) \times (k+1)$ array Counts of response frequencies, where Counts$[a][b][c]$ is the number of tasks for which worker 1 responded with $r_a$, worker 2 responded with $r_b$, and worker 3 responded with $r_c$. Here, each of $a$, $b$, $c$ can either be 0 (if that worker did not attempt the task) or $i \in \{1, 2, ..k\}$ (if the worker's response was $r_i$). For instance, Counts$[0][3][1]$ is the number of times worker $w_1$ did not attempt the task, worker $w_2$ responded with $r_3$, and worker $w_3$ responded with $r_1$.

To begin with, we describe the method ProbEstimate, which takes in the 3-dimensional array of counts as input and returns a point estimate of the worker response probabilities (actually it returns an estimate of $S_D^{\frac{1}{2}}P_i$, which can be used to deduce the response probabilities). We will later make multiple calls to method ProbEstimate to obtain confidence intervals for response probabilities. In Step 1 in Algorithm A3, we compute the number of tasks attempted by all three workers (we call this $n_{1,2,3}$) and the number of tasks attempted by each pair of workers $w_i$ and $w_j$ (we call this $n_{i,j}$). Tasks attempted by fewer than two workers will not be used by our method.

In Step 2, we compute response frequency matrices. These matrices give us the probability of getting each pair of responses from each pair of workers. Specifically, $R_{i_1,i_2}(j_1,j_2)$, the element in the $j_1^{th}$ row and $j_2^{th}$ column of $R_{i_1,i_2}$ gives us the probability that worker $w_{i_1}$ will respond with $r_{j_1}$ and worker $w_{i_2}$ will respond with $r_{j_2}$ to a random task they've both attempted. To estimate the probability $R_{i_1,i_2}(j_1,j_2)$, we take the number of tasks in which worker $w_{i_1}$ responded with $r_{j_1}$ and $w_{i_2}$ responded with $r_{j_2}$ and divide it by the total number of tasks attempted by both workers. For instance, if worker $w_1$ and $w_2$ attempted 100 tasks in common, and if $w_1$ responded with $r_1$ and $w_3$ with $r_2$ on 30 of those tasks, then our estimate for $R_{1,3}(1,2)$ would be $\frac{30}{100}$. Lemma 6 tells us the relation between the response frequency matrix and the response probability matrices.

**Lemma 6.** Let $R_{i_1,i_2}$ be the response frequency matrix so that $R_{i_1,i_2}(j_1,j_2)$ gives the probability of worker $w_{i_1}$ responding with $r_{j_1}$ and $w_{i_2}$ responding with $r_{j_2}$. Then $R_{i_1,i_2} = P_{i_1}^T S_D P_{i_2}$, where $P_i$ is the response probability matrix of $w_i$ and $S_D$ is the diagonal form of the selectivity matrix □

We get three such equations by setting $(i_1,i_2)$ to $(1,2), (3,2)$ and $(3,1)$. Solving them gives us the result of Lemma 7, that is, $S_D^{\frac{1}{2}}P_1 = UED^{\frac{1}{2}}E^{-1}$, where $U$ is some unitary matrix and $EDE^{-1}$ is the eigen-decomposition of $R_{1,2} \times R_{3,2}^{-1} \times R_{3,1}$.

**Lemma 7.** Let $EDE^{-1}$ be the eigen-decomposition of $R_{1,2} \times R_{3,2}^{-1} \times R_{3,1}$, where $R_{1,2}, R_{3,2}, R_{3,1}$ are response frequency matrices. Then

$$R_{1,2} \times R_{3,2}^{-1} \times R_{3,1} = P_1^T S_D P_1 = (S_D^{\frac{1}{2}}P_1)^T(S_D^{\frac{1}{2}}P_1)$$

and there exists a unitary matrix $U$ such that we have $S_D^{\frac{1}{2}}P_1 = UED^{\frac{1}{2}}E^{-1}$ □

Thus in Step 3, we take the eigen-decomposition $EDE^{-1}$ of $R_{1,2} \times R_{3,2}^{-1} \times R_{3,1}$, where each column of $E$ is an eigenvector and $D$ is a diagonal matrix whose diagonal elements are the eigenvalues. Then in Step 4, we compute $U_1 = ED^{\frac{1}{2}}E^{-1}$, $U_2 = (U_1^T)^{-1}R_{1,2}$ and $U_3 = (U_1^T)^{-1}R_{1,3}$. By Lemma 6 and 7, there exists a unitary matrix $U$ such that $S_D^{\frac{1}{2}}P_1 = UU_1$, $S_D^{\frac{1}{2}}P_2 = UU_2$ and $S_D^{\frac{1}{2}}P_3 = UU_3$. Thus it remains to find $U$.

In Step 6.$a$, we compute the number of tasks where worker $w_3$ responded with $r_{j_3}$, denoted $n_{j_3}$. Then in Step 6.$b$, we compute the conditional response frequency matrix $R_{1,2,3,j_3}$. This matrix is similar to the matrix $R_{1,2}$ from Step 2, but only considers tasks that $w_3$ has attempted and responded to with $r_{j_3}$. That, $R_{1,2,3,j_3}(j_1,j_2)$ is the probability of $w_1$ responding with $r_{j_1}$ and $w_2$ responding with $r_{j_2}$ given that worker $w_3$ has responded with $r_{j_3}$. We estimate this by counting the number of times $w_1$ responded with $r_{j_1}$, $w_2$ responded with $r_{j_2}$ and $w_3$ responded with $r_{j_3}$, and dividing by the $n_{j_3}$. For instance, if worker $w_3$ responded to 40 tasks with $r_2$, and among those 40 there were 10 tasks in which $w_1$ responded with $r_4$ and $w_2$ with $r_3$, then $R_{1,2,3,r_2}(r_4,r_3) = \frac{10}{40}$. Next, we use Lemma 8, which tells us the relation between the conditional response frequency matrix and the unitary matrix $U$.

**Lemma 8.** Let $R_{1,2,3,j_3}$ be the conditional response frequency matrix so that $R_{1,2,3,j_3}(j_1,j_2)$ gives the probability of worker $w_1$ responding with $r_{j_1}$ and worker $w_2$ responding with $r_{j_2}$ given that worker $w_3$ has responded with $r_{j_3}$. Then

$$R_{i_1,i_2} = (S_D^{\frac{1}{2}}P_{i_1})^T W_{3,j_3} S_D^{\frac{1}{2}}P_{i_2}$$

where $P_i$ is the response probability matrix of worker $w_i$ and $S_D$ is the diagonal form of the selectivity matrix, and $W_{3,j_3}$ is the diagonal form of the $j_3^{th}$ column of $P_3$. □

---

**Begin function ProbEstimate**

ProbEstimate $[V_1, V_2, V_3] \leftarrow$ ProbEstimate(Counts)

1) Find $n_{1,2,3}$, the number of tasks attempted by all three workers, and $n_{i,j}$, the number of tasks attempted by worker $w_i$ and $w_j$ only, for $(i,j) \in \{(1,2),(2,3),(3,1)\}$.
2) Find the response frequency matrices $R_{1,2}, R_{2,3}$, and $R_{3,1}$ as follows. For each $j_1 \in \{1,2,...k\}, j_2 \in \{1,2,...k\}$ :
   a) $R_{1,2}(j_1,j_2) = \frac{\sum_{j_3=0}^{k} \text{Counts}[j_1][j_2][j_3]}{n_{1,2,3}+n_{1,2}}$.
   b) $R_{2,3}(j_1,j_2) = \frac{\sum_{j_3=0}^{k} \text{Counts}[j_3][j_1][j_2]}{n_{1,2,3}+n_{2,3}}$.
   c) $R_{3,1}(j_1,j_2) = \frac{\sum_{j_3=0}^{k} \text{Counts}[j_2][j_3][j_1]}{n_{1,2,3}+n_{3,1}}$.
   Let $R_{2,1} = R_{1,2}^T$, $R_{3,2} = R_{2,3}^T$ and $R_{1,3} = R_{3,1}^T$
3) Let $EDE^{-1}$ be the eigenvalue decomposition of $R_{1,2} \times R_{3,2}^{-1} \times R_{3,1}$, where each column of $E$ is an eigenvector, and $D$ is a diagonal matrix whose diagonal elements are eigenvalues.
4) Let $U_1 = ED^{\frac{1}{2}}E^{-1}$, where $D^{\frac{1}{2}}$ is the element-wise square root of $D$. Let $U_2 = (U_1^T)^{-1}R_{1,2}$ and $U_3 = (U_1^T)^{-1}R_{1,3}$.
5) Initialize $V_1$ to the $k \times k$ matrix with all zeros.
6) For each $j_3 \in \{1,2...k\}$, do:
   a) $n_{j_3} = \sum_{j_1=1}^{k}\sum_{j_2=1}^{k}$ Counts $[j_1][j_2][j_3]$.
   b) Find the conditional response frequency matrix $R_{1,2,3,j_3}$ as follows. For each $j_1 \in \{1,2,...k\}, j_2 \in \{1,2,...k\}$ : $R_{1,2,3,j_3}(j_1,j_2) = \frac{\text{Counts}[j_1][j_2][j_3]}{n_{j_3}}$.
   c) Let $U^{-1}WU$ be the eigenvalue decomposition of $(U_1^T)^{-1}R_{1,2,3,j_3}U_2^{-1}$, where columns of $U^{-1}$ are eigenvectors and $W$ is a diagonal matrix whose diagonal values are eigenvalues.
   d) Let $V_{1,j_3} = UU_1$. For each $j \in \{1,2...k\}$ : Find the largest element in the $j^{th}$ row of $V_{1,j_3}$. Let the column index of that element be $j'$. Then swap the $j^{th}$ and $j'^{th}$ rows of $V_{1,j_3}$.
   e) $V_1 = V_1 + \frac{V_{1,j_3}}{k}$.
7) Let $V_2 = (V_1^T)^{-1}R_{1,2}$ and $V_3 = (V_1^T)^{-1}R_{1,3}$ .
8) Return $[V_1, V_2, V_3]$.

---

We find the eigen-decomposition $U^{-1}WU$ of $(U_1^T)^{-1}R_{1,2,3,j_3}U_2^{-1}$, where each column of $U^{-1}$ is an eigenvector and $W$ is a diagonal matrix with eigenvalues as its elements. $U$ is an estimate for the $U$ from $S^{\frac{1}{2}}P_i = UU_i$ (Lemma 7), but with some rows possibly swapped with other rows. To get the correct order of rows, we multiply $U$ by $U_1$ to get an estimate for $V_1 = S^{\frac{1}{2}}P_1$. Recall that for every worker $w_i$, the diagonal element in each row of the response probability matrix $P_i$ is the largest element in that row. We use this to bring rows of $V_1$ to their correct positions in Step $6.d$ by swapping rows to make the diagonal element the largest in its row. This gives us an estimate of $V_1 = S^{\frac{1}{2}}P_1$.

Our final estimate for $V_1$ is the average of the estimates obtained from each $j_3$. In Step 7, we estimate $V_2$ as $(V_1^T)^{-1}R_{1,2}$ and $V_3$ as $(V_1^T)^{-1}R_{1,3}$. Finally, we return the values of $V_1$, $V_2$, $V_3$ in Step 8.

The method ProbEstimate only gives us a point estimate of the worker response probabilities. However, our aim is to find not only point estimates, but also confidence intervals for the response probabilities. Our method to get confidence intervals is also in Algorithm A3. We use Theorem 1 to get the required confidence intervals. We already have a function ProbEstimate that maps Counts to response probabilities. This function will act as our function $f$ from Theorem 1. In addition, we need to know the covariance between each pair of elements in Counts. This is done in Step 4, using Lemma 9.

**Lemma 9.** Let Counts $[i_1][i_2][i_3]$ denote the number of times $w_1$ responded with $r_{i_1}$, $w_2$ responded with $r_{i_2}$, and $w_3$ responded with $r_{i_3}$, where a response of $r_0$ indicates that the worker did not attempt the task. Then the covariance between Counts $[i_1][i_2][i_3]$ and Counts $[j_1][j_2][j_3]$ is obtained as follows:

1) **if** $(i_t = 0 \wedge j_t \neq 0) \vee (i_t \neq 0 \wedge j_t = 0)$ for some $t \in \{1,2,3\}$ :

$$\text{Cov}(\text{Counts}[i_1][i_2][i_3], \text{Counts}[j_1][j_2][j_3]) = 0$$

2) **else if** $i_t = j_t$ for every $t \in \{1,2,3\}$ :

$$\text{Cov}(\text{Counts}[i_1][i_2][i_3], \text{Counts}[j_1][j_2][j_3])$$
$$\approx \frac{\text{Counts}[i_1][i_2][i_3](n - \text{Counts}[i_1][i_2][i_3])}{n}$$

where $n$ is the number of tasks attempted by exactly the set of workers corresponding to responses $r_{i_1}, r_{i_2}, r_{i_3}$ (for example, if $i_2 = 0$ and $i_1, i_3 \neq 0$, then $n$ is the number of tasks attempted by workers $w_1$ and $w_3$ only).

3) **else** $\text{Cov}(\text{Counts}[i_1][i_2][i_3], \text{Counts}[j_1][j_2][j_3])$
$$\approx -\frac{\text{Counts}[i_1][i_2][i_3]\text{Counts}[j_1][j_2][j_3]}{n}$$

where $n$ is the number of tasks attempted by exactly the set of workers corresponding to responses $r_{i_1}, r_{i_2}, r_{i_3}$. □

Finally, we need to know the derivative of each element of each $S_D^{\frac{1}{2}}P_i$ with respect to each element in Counts, to get a linear approximation of function ProbEstimate to use Theorem 1. We compute these derivatives numerically.

To start with, we fix a small $\epsilon$ in Step 5. Then in Step 6, for each element of Counts, we increment it by $\epsilon$ (Step $6.a$) and run ProbEstimate on the modified Counts array, to get $[V_1', V_2', V_3']$ in Step $6.b$. We then decrease that element by $2\epsilon$ in Step $6.c$(thus taking it $\epsilon$ below the original value) and run ProbEstimate to get $[V_1'', V_2'', V_3'']$ in Step $6.d$. Then we set the element back to its original value by adding $\epsilon$.

The derivative of each response probability with respect to an element $e$ is given by its value for $e + \epsilon$ minus its value for $e - \epsilon$, divided by $2\epsilon$. This derivative is found in Step $6.f.i$.

Finally, using the derivatives and covariances computed in previous steps, we apply Theorem 1 to get confidence intervals for response probabilities. (We actually get confidence intervals for $S_D^{\frac{1}{2}}P_i$, and we can normalize the mean of intervals in each row to get confidence intervals for elements of $P_i$).

---

**Algorithm A3:**

1) Find $n_{1,2,3}$, the number of tasks attempted by all three workers, and $n_{i,j}$, the number of tasks attempted by worker $w_i$ and $w_j$ only, for $(i,j) \in \{(1,2),(2,3),(3,1)\}$.
2) Find the response frequency matrices $R_{1,2}$, $R_{2,3}$, and $R_{3,1}$ as follows. For each $j_1 \in \{1,2,...k\}, j_2 \in \{1,2,...k\}$ :
   a) $R_{1,2}(j_1, j_2) = \frac{\sum_{j_3=0}^{k} \text{Counts}[j_1][j_2][j_3]}{n_{1,2,3}+n_{1,2}}$.
   b) $R_{2,3}(j_1, j_2) = \frac{\sum_{j_3=0}^{k} \text{Counts}[j_3][j_1][j_2]}{n_{1,2,3}+n_{2,3}}$.
   c) $R_{3,1}(j_1, j_2) = \frac{\sum_{j_3=0}^{k} \text{Counts}[j_2][j_3][j_1]}{n_{1,2,3}+n_{3,1}}$.
   Let $R_{2,1} = R_{1,2}^T$, $R_{3,2} = R_{2,3}^T$ and $R_{1,3} = R_{3,1}^T$
3) $[V_1, V_2, V_3]$ = ProbEstimate(Counts).
4) For each $i_1, i_2, i_3, j_1, j_2, j_3$ from 1 to $k$, find the covariance between Counts $[i_1][i_2][i_3]$ and Counts $[j_1][j_2][j_3]$ using Lemma 9.
5) We now find the derivative of each element of each $P_i$ with respect to each element of Counts. Fix a small $\epsilon$, say 0.01.
6) For each $j_1, j_2, j_3$ from 1 to $k$:
   a) Counts $[j_1][j_2][j_3]$ = Counts $[j_1][j_2][j_3] + \epsilon$.
   b) $[V_1', V_2', V_3']$ = ProbEstimate(Counts).
   c) Counts $[j_1][j_2][j_3]$ = Counts $[j_1][j_2][j_3] - 2\epsilon$.
   d) $[V_1'', V_2'', V_3'']$ = ProbEstimate(Counts).
   e) Counts $[j_1][j_2][j_3]$ = Counts $[j_1][j_2][j_3] + \epsilon$.
   f) For each $i \in \{1,2,3\}$, and each $i_1, i_2$ from 1 to $k$:
      $\frac{\partial P_i(i_1, i_2)}{\partial \text{Counts}[j_1][j_2][j_3]} \approx \frac{V_i'(i_1, i_2) - V_i''(i_1, i_2)}{2\epsilon}$.
7) Using Theorem 1, the covariances between each pair of Counts, and the derivatives computed in the previous step, find confidence intervals for $P_i(i_1, i_2)$ for each $i \in \{1,2,3\}$ and each $i_1, i_2$ from 1 to $k$.

---

If the total number of tasks is $n$, then computational complexity per worker evaluated is $O(k^6 + nk^3)$. While this grows very fast with $k$, typical crowdsourcing applications involve very small values of $k$ and larger values of $n$, and therefore, the computational cost is not very significant, especially in comparison with the latency of crowdsourcing.

*B. Experiments on Synthetic Data*

We now experimentally evaluate our confidence intervals to study their performance. We perform experiments on synthetic data in this section, on real data in the next section.

For the synthetic data experiments, there are a potentially huge number of scenarios to consider. We conducted a large number of experiments, but we only have space to summarize a couple of representative experiments. We consider arities $k = 2, 3, 4$. For each arity, we consider three possible worker response probability matrices, and each worker is assigned one of the matrices with equal probability independently of other workers. The response probabilities in the matrices are chosen arbitrarily, but we observe similar results for other values of the response probabilities. The matrices are as follows:

Arity 2: $\begin{bmatrix} 0.9 & 0.1 \\ 0.2 & 0.8 \end{bmatrix}, \begin{bmatrix} 0.8 & 0.2 \\ 0.1 & 0.9 \end{bmatrix}, \begin{bmatrix} 0.9 & 0.1 \\ 0.1 & 0.9 \end{bmatrix}$

Arity 3: $\begin{bmatrix} 0.6 & 0.3 & 0.1 \\ 0.1 & 0.6 & 0.3 \\ 0.3 & 0.1 & 0.6 \end{bmatrix}, \begin{bmatrix} 0.8 & 0.1 & 0.1 \\ 0.2 & 0.8 & 0.0 \\ 0.0 & 0.2 & 0.8 \end{bmatrix}, \begin{bmatrix} 0.9 & 0.0 & 0.1 \\ 0.1 & 0.9 & 0.0 \\ 0.0 & 0.2 & 0.8 \end{bmatrix}$

Arity 4:

$\begin{bmatrix} 0.7 & 0.1 & 0.1 & 0.1 \\ 0.1 & 0.6 & 0.2 & 0.1 \\ 0.0 & 0.1 & 0.8 & 0.1 \\ 0.2 & 0.1 & 0.0 & 0.7 \end{bmatrix}, \begin{bmatrix} 0.8 & 0.1 & 0.0 & 0.1 \\ 0.1 & 0.8 & 0.0 & 0.2 \\ 0.1 & 0.1 & 0.7 & 0.1 \\ 0.0 & 0.1 & 0.2 & 0.7 \end{bmatrix}, \begin{bmatrix} 0.6 & 0.1 & 0.2 & 0.1 \\ 0.0 & 0.7 & 0.1 & 0.2 \\ 0.1 & 0.0 & 0.9 & 0.0 \\ 0.2 & 0.0 & 0.0 & 0.8 \end{bmatrix}$

*1) Confidence vs. Accuracy:* In this experiment, we fix a number of tasks $n$, arity $k$ and a confidence level $c$. The true response for each task is assumed to be one of $r_1, r_2, ... r_k$, each with equal probability $\frac{1}{k}$. We have three workers, and each of them attempts all $n$ tasks (the result of the experiment is similar even if each worker attempts only a fraction of the tasks). This gives us our data. We run our 3-worker $k$-ary non-regular method on this data to get $c$-confidence intervals for each element of the response probability matrix of each worker. For each interval computed, we check to see if the true value of the worker response probability lies within the confidence interval. We repeat this experiment 500 times, with different sets of workers, and compute the interval-accuracy i.e., the number of confidence intervals that contained the true response probability, divided by the total number of intervals. This interval-accuracy is computed for every $c \in \{0.05, 0.1, ... 0.95\}$ and plotted against $c$ in Figure 5(a). We make such a plot for number of tasks $n = 100, 1000$ and for arity $k = 2, 3, 4$.

The solid line $x = y$ in Figure 5(a) represents the ideal accuracy level. The plot shows that when the number of tasks is small, and arity $k > 2$, the interval-accuracy is somewhat higher than the ideal accuracy level, suggesting that our method is extra conservative when the amount of data is small relative to the arity. On the other hand, when the amount of data is larger relative to the arity (for arity 2 or $n = 1000$), our method achieves almost exactly the level of accuracy desired.

*2) Density and Arity vs. Interval Size:* In this experiment, we study the relation between the density of data (the fraction of worker-task pairs in which the worker actually attempted the task) and arity on one hand and interval size on the other. We fix the number of tasks $n$ to be 500 and confidence level $c$ to be 0.8. The true response for each task is equally likely to be any of $r_1, r_2, ... r_k$. We then choose an arity $k$ and density value $d$. There are three workers, each of whom attempts every task with probability $d$, independently of which other tasks the worker has attempted or which tasks other workers have attempted. We then run our 3-worker $k$-ary non-regular method on the resulting data to get $c$-confidence intervals for each response probability of each worker. We repeat this experiment 500 times and compute the average interval size of all confidence intervals generated. This average interval size is computed for $d \in \{0.5, 0.55...0.95\}$ and plotted against $d$ in Figure 5(b). We make such a plot for $k = 2, 3, 4$.

Figure 5(b) shows that average interval size increases as density decreases, for all values of $k$. This is expected, as decreasing density reduces the amount of data available to our method. Moreover, the plot shows that increasing arity significantly increases average interval size. This is because the number of variables we have to estimate is directly proportional to the square of arity $k$. Thus as we increase arity while keeping number of tasks constant, the amount of data available per variable goes down, leading to an increase in average interval size.

*C. Experiments on Real Data*

We now evaluate our confidence intervals on real data to test their performance in situations where our assumptions may not hold. We use three different datasets, which we call MOOC(Massive Open Online Course), WSD(Word Sense), and WS(Word Similarity). The datasets are described in the next subsection. Gold standard responses to all tasks are known in WSD and WS, while in MOOC, we only experiment using tasks for which gold standard responses are known. Since we do not know the true response probabilities of workers, we compute the fraction of times a worker $w_i$ gave response $r_{j_2}$ when the true response was $r_{j_1}$ and use that fraction as a proxy for the true response probability $P_i(j_1, j_2)$.

*1) Data and Settings:* The MOOC dataset is a set of peer evaluations from a Massive Open Online Course [31]. Students were asked to grade their peers' assignments, providing a grade from 0 to 5. Some of the assignments were also graded by course assistants, and these grades are treated as gold standard grades. We only experiment using assignments with gold standard grades. This gives us a dataset with 6-ary tasks. Since the amount of data is too low relative to the arity 6, we turn this into a 3-ary task by mapping each grade $g$ to $\lfloor \frac{g}{2} \rfloor$.

The other two datasets, WS and WSD, are from [30]. In WS, workers were given a pair of words and were asked to give a similarity rating to the words from 0 to 10. The dataset is 11-ary, but is so sparse that no triple of workers had more than 30 tasks in common. Hence we reduce the arity of the task to 2 by replacing each rating $g$ by $\lfloor \frac{g}{6} \rfloor$.

In WSD, workers were given a sentence, with one word highlighted. They were also given three different choices as to what the word meant in context of the sentence. The worker had to respond with 1, 2, or 3 based on the index of the most appropriate meaning. Although the data is actually 3-ary, there are almost no tasks whose true response is 2, making the data practically binary. Our technique doesn't work on the dataset with arity set to 3 because one of the matrix rows has only zeros, making it non-invertible. To avoid this, we reduce the dataset to a binary dataset, mapping responses 2 and 3 to 2.

*2) Confidence vs. Accuracy:* We start by fixing a confidence level $c$. We wish to study the accuracy of our $c$-confidence intervals on the datasets. Our datasets are fairly sparse, and many pairs of workers have almost no tasks in common. Hence, we wish to restrict attention to sets of workers that have a reasonable number of tasks in common. Therefore for each dataset, we choose a threshold $t$ such that there are at least 50 triples of workers that have attempted $t$ tasks in common. We choose $t = 60$ for MOOC, $t = 100$ for WSD, and $t = 30$ for WS. The datasets all have several workers, while our method only applies to 3 workers, so we consider a set of 3 workers at a time and apply our method to their responses. To do this, we choose a random triple of workers that has attempted at least $t$ tasks in common and run the 3-worker $k$-ary non-regular method on their responses to get confidence intervals for their response probabilities. For each interval, we check to see if the true response probability lies within the interval. We do this for 50 triples of workers and compute the interval-accuracy i.e., the number of intervals

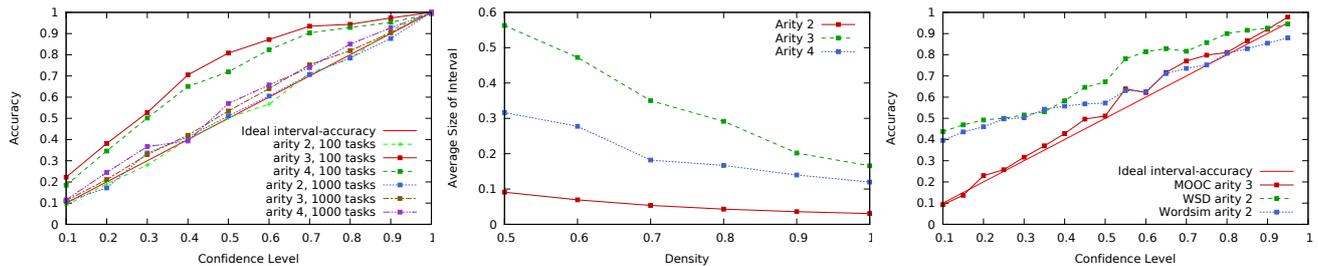

Fig. 5. (a)Accuracy of confidence interval vs confidence level (b) Average Size of confidence interval vs density (c)Accuracy of confidence interval vs confidence level

that contain the true response probability, divided by the total number of intervals. This interval-accuracy is found for all values of $c \in \{0.05, 0.1, ...0.95\}$ and plotted against $c$ in Figure 5(c). This is done for all the datasets.

The solid line $x = y$ in Figure 5(c) represents the ideal accuracy level. The plot shows that for MOOC, the interval accuracy is almost equal to the ideal accuracy level. For the other two datasets, our intervals have somewhat higher accuracy than required for low confidence levels, but approach the line as confidence level rises. Since higher confidence levels tend to be used in practice, our intervals perform well in most practical situations.

Our method was based on multiple assumptions, such as the local linearity of function ProbEstimate for use in Theorem 1 or tasks having uniform difficulty. These assumptions may not hold fully in real data, but Figure 5(c) shows that our intervals still perform well in such cases. Thus our method not only works on $k$-ary non-regular data, but is also robust to small violations of our assumptions and produces reliable confidence intervals in such cases.

## V. Conclusion

In this paper, we introduced techniques for finding confidence intervals for worker error rates in very general conditions. Confidence intervals tell us how reliable our estimates are and they are very important when deciding which workers to retain and which workers to fire. Using confidence intervals allows us to end up with a good set of workers faster than we could by using mean error estimates [2], yielding improved quality crowdsourced results. Our techniques provide confidence intervals for tasks having two or more possible answers, even when data are not regular, and also take into account possible biases of workers toward some answers. As described, our methods work on the entire dataset in a one-time fashion, but they can be easily modified to be incremental, to keep efficiently updating worker error rates as more tasks get done. Our experimental results show that our techniques yield accurate and useful confidence intervals, even in cases where the few assumptions we make do not hold.